\documentclass[aps,prd,nofootinbib,10pt]{revtex4-1}
\usepackage[colorlinks=true,linkcolor=black,citecolor=blue,urlcolor=black]{hyperref}
\usepackage{natbib}
\usepackage[T1]{fontenc}
\usepackage{graphicx}
\usepackage{empheq}
\usepackage{slashed}
\usepackage{amsmath,float}
\usepackage{amsfonts}
\usepackage{amssymb}
\usepackage{xcolor}
\usepackage{framed}
\usepackage[toc,page]{appendix}
\usepackage{braket}
\usepackage{caption}
\usepackage{color}
\usepackage[normalem]{ulem}

\def\beq{\begin{equation}}
\def\eeq{\end{equation}}
\def\barr{\begin{eqnarray}}
\def\earr{\end{eqnarray}}
\def\dis{\displaystyle}

\def\gev{\, {\rm GeV}}
\def\tev{\, {\rm TeV}}

\begin{document}
\title{Discussing 125\,GeV and 95\,GeV excess in Light Radion Model}

\author{Divya Sachdeva}
\email{divyasachdeva951@gmail.com}
\author{Soumya Sadhukhan}
\email{physicsoumya@gmail.com}
\affiliation{Department of Physics and Astrophysics, University of Delhi, Delhi 110 007, India}

\begin{abstract}
 Even if the LHC observations are consistent with the Standard model (SM), 
 current LHC results are not precise enough to rule out the presence of new physics. Taking a contrarian view of the SM Higgs fandom,
we look out for a more suitable candidate for the 125\,GeV boson observed at the LHC. At the same time, a recent
result from CMS hints towards an excess near 95\,GeV \ in the diphoton ($\gamma \gamma$) channel. Given these aspects, we revisit the
Higgs-radion mixing model to explore the viability of the radion mixed Higgs to be the 125\,GeV boson along with the
presence of a light radion (to be precise Higgs mixed radion) that can show up in future experiments in the $\gamma \gamma$ channel. We find that the
mixed radion-Higgs scenario gives a better fit than the SM, with the radion mixed Higgs as a more suitable 125\,GeV 
scalar candidate. It also gives rise to a diphoton excess from the light radion, consistent with the LHC observations.
\end{abstract}

\maketitle

\section{Introduction}
The discovery of a boson of mass 125\,GeV at the LHC validates the standard model to be the
most predictive model of particle physics. As LHC probes the boson further, it turns out that
its interactions match the Standard Model (SM) Higgs boson, albeit with minor exceptions. 
Still, the debate if the SM is the ultimate theory of the particle world is far from over and LHC
results with large uncertainties can disclose new possibilities. The current measurement of the Higgs
signal strengths~\cite{ATLAS:2018doi,
Sirunyan:2018koj,Cheung:2018ave} still allow for small yet significant deviation from the SM values. Therefore, it is
worthwhile to look for a more suitable Higgs candidate than the SM Higgs, which can be explored in 
different beyond the Standard model (BSM) scenarios. As the LHC observations indicate presence of a
scalar that mimics the SM Higgs, only a model with minimal modification in the scalar sector is likely
to accommodate a more interesting alternative. As we invoke BSM scenarios containing a Higgs like scalar, it would be 
interesting if these can also address the issues plaguing SM viz the resolution of gauge hierarchy problem, an explanation 
of the baryon asymmetry, offering a dark matter candidate etc.

 Even after the 8\,TeV and 13\,TeV run of the LHC with increasing luminosities, any trace of new 
 physics at the TeV scale that can potentially fix the gauge hierarchy problem, is yet to be discovered. If this is taken to imply that
 new physics exists only at energies higher than the TeV scale, a little fine tuning is automatically introduced. It can be avoided if
 the new physics search is concentrated around the Higgs mass scale. 
Furthermore, 
it is quite conceivable that the new physics can be hidden at a lower scale i.e. lighter than the Higgs, instead of always at 
a scale higher ($\sim$TeV).
Specially in the hadron colliders 
like the LHC, probing the lighter spectrum is not that efficient due to enormous QCD background. Therefore, 
it is pertinent to consider innovative model construction aided by enhanced signal strength where BSM physics 
can be probed at the sub 100 GeV scale.
%
Another recent motivation that prods us to probe at lower scales is the variety of mild excesses we have 
observed over the years~\cite{Barate:2003sz,Sirunyan:2017nvi,Sirunyan:2018wim}. Amongst these observations, the most recent result from CMS~\cite{Sirunyan:2018aui} shows a small 
excess near 95 GeV in the diphoton channel. All of these observation compel us to look out for the new 
physics models at these scales carefully, especially the one where diphoton resonance can be an important
feature. 

To arrange for a better 125\,GeV Higgs-like candidate along with a light spectrum, extended scalar sector BSM scenarios 
can be delved into. While a new scalar discovery in future experiments will compel us to explore beyond minimal Higgs
sector of the SM, the 125\,GeV particle as the only observed scalar can also have an underlying extended scalar sector. The simplest
extension of the SM scalar sector is to add a singlet scalar. However, as it mixes minimally with the Higgs, 
it is unable to give rise to an excess in the diphoton channel over other channels and is, therefore, of no major interest for this work. Higgs 
sector extended with another $SU(2)_L$ doublet, motivated by the supersymmetric and grand unified theories, also 
severely constrains presence of a light scalar due to sum rule of scalar couplings to fermions and gauge bosons,
as discussed in Ref.~\cite{Choudhury:2003ut}. In this regard, the radion, a scalar introduced in extra dimensional model to stabilize the geometry, 
is discussed. Being the Goldstone boson of the scale invariance breaking, it has trace anomaly-induced couplings to the
massless bosons (photons as well as gluons) and consequently allows for a distinct possibility of a non-trivial diphoton
decay. Contrary to the other BSM scalars, it can be really light with mass $\sim 100 \gev$ as had been shown in Randall Sundrum (RS) 
model~\cite{Goldberger:1999uk,Kribs:2006mq}. In addition, it can mix to the Higgs boson via curvature scalar mixing and this can alter the Higgs couplings 
significantly. Therefore it behoves us to probe radion-Higgs mixing scenario to assess the viability of a non-standard Higgs 
as the 125\,GeV scalar at the LHC.
The radion Higgs mixing as has been explored later in our work can significantly modify the couplings of both the scalar mass-eigenstate.
More specifically, the Higgs gluon gluon coupling is enhanced due to the contribution from trace anomalous part of radion
gluon gluon vertex, which is found to help in explaining the signal strength of the Higgs signal better than the SM Higgs itself. 
The significant parameter region where the radion mixed Higgs is a more suitable candidate for the 125 GeV boson instead of the SM Higgs hitherto pitched 
so aggressively, is presented. 
If we explore the lighter than 125\,GeV side of the spectrum in the context of a diphoton excess, the RS model radion can be a 
suitable candidate. The radion gamma gamma vertex is modified compared to that of a SM like Higgs due to its trace anomaly part and this can 
increase its branching ratio to the diphoton channel with respect to the other fermionic and gauge boson channels. Therefore it is
worthwhile to propose radion as a natural candidate that can potentially give rise to the diphoton excess.

While some of the observations in this paper have already been noted in the previous works~\cite{Csaki:2000zn,Desai:2013pga,Chakraborty:2017lxp} 
and the diphoton excess due to light scalar has also been discussed in several papers~\cite{Vega:2018ddp,Liu:2018xsw,Kim:2018mks,Han:2018bni,Biekotter:2017xmf,
Haisch:2017gql,Cacciapaglia:2017iws,Fox:2017uwr,Richard:2017kot,Tao:2018zkx,Torre:2018jnf,LiuLiJia:2019kye,Biekotter:2019mib,Choi:2019yrv},
the data used are current, leading to new bounds, and, we show that the extension of SM by such a scalar can
actually give better fit to the Higgs signal measurment. We begin our discussion with a short review on radion in RS model so that this paper can be read as far
as possible independently of the preceding literature. Next, in section III, we discuss phenomenological 
and theoretical constraints on parameter space of radion and then show, in Section IV how the new scalar 
may explain the recent CMS excess near 95 GeV, as well as can give the better fit to the Higgs signal measurement.
We conclude in Section V.

\section{Model Description}
We first introduce the Minimal RS model and show how the radion can appear here, outlining its interaction with the SM 
particles. Then we analyse radion Higgs mixing through scalar-curvature interaction, listing the modified 
couplings for both the scalars. 
\subsection*{Minimal RS model}
In the minimal version of RS model, an extra warped dimension of radius $r_c$ is compactified 
down to a $S_1/Z_2$ orbifold. The orbifolding is applied with a
pair of 3-branes at the fixed points $x_4 = 0$ and $x_4=r_c \pi$.
The brane at $x_4=0$, where gravity peaks, is called the Planck (hidden)
brane, while the brane where SM fields are confined is called the $\tev$
(visible) brane. Note that there are many other version of the model
where fields other than graviton are allowed to propagate in
the bulk
however, we limit ourselves to the minimal case. The action for this set 
up is given by\cite{Randall:1999ee}
\begin {eqnarray}
 S & = & S_{gravity} + S_{v} + S_{h} \nonumber \\ 
 S_{gravity} & = & \int d^5x \sqrt{-g} \{2M_5^3R-\Lambda\} \nonumber \\ 
 S_{v} & = & \int d^4 x \sqrt{-g_{v}} \{{\cal{L}}_{v} - V_{v}\} \nonumber \\ 
 S_{h} & = & \int d^4 x \sqrt{-g_{h}} \{{\cal{L}}_{h} - V_{h}\}  
\end {eqnarray}
where $g$ is the determinant of the five dimensional
metric $g_{MN}(x_\mu,x_4)$, the greek indices being representation of
(1+3) dimensional coordinates on the visible (hidden) brane and $M_{5}$ is the
5-dimensional Planck mass and $\Lambda$ is the bulk cosmological
constant.  $V_{v}$ and $V_{h}$ are the brane tensions of visible and
hidden branes respectively.

After solving Einstein's equations, the metric has form
\begin{equation}
ds^2 = e^{-2k|x_4|} \eta_{\mu\nu} dx^\mu dx^\nu + dx_4^2
\end{equation}
where $k = \dis\sqrt{\frac{-\Lambda}{24 M_{5}^3}}$, $V_{h} = -V_{v} = 24M_5^3k$, 
$|x_4|=r_c\phi$.
$M_5$ is related to the four dimensional Planck mass, $M_{Pl}$ as
\begin{equation}
 M^2_{Pl} = \frac{M^3_{5}}{k}[1 - e^{-2k r_c\pi}]
\end{equation}
A field with mass $m$ propagating on the visible brane in the 5-dimensional theory
generates an effective mass $m_{eff} = m e^{-kr_c\pi}$ in the 4D effective theory. To solve 
the hierarchy problem, one needs $k \, r_c\sim 12$. With this value, the Planck scale
is reduced to the weak scale. However, for the 
background metric solution discussed above, any value of the radius $r_c$ is equally possible.
Therefore a mechanism is needed to fix it uniquely with the desired value so that the EW hierarchy
can be explained. One of the mechanism~\cite{Goldberger:1999uk,Cline:2000xn,Choudhury:2000wc,Anand:2014vqa}
that addresses this issue was given by Goldberger and Wise(GW)~\cite{Goldberger:1999uk}.

In GW mechanism, $r_c$ is considered as the vev of a modulus field $F(x)$ that quantifies the
fluctuation about the radius:
\[ds^2 = \dis e^{-2k|\theta|F(x)} g_{\mu\nu} dx^\mu dx^\nu - F^2(x)d\theta^2\]
Upon reducing the 5D Einstein Hilbert action for this metric, the following effective action 
is obtained:
\begin{eqnarray}
\mathcal{S}&=& \dis M_5^3\int d^4x d\theta \sqrt{-g} e^{-2k\theta F(x)}\left(6\,k\,|\theta|\partial_\mu F(x)\partial^\mu F(x)\right. \nonumber\\
&-& \left. 6 k^2|\theta|^2F\partial_\mu F(x)\partial^\mu F(x) + F(x)\,R\right)
\end{eqnarray}
where $R$ is 4D Ricci scalar. After we integrate out $\theta$, we get the following 4D action for $\Phi\,=\,\dis \sqrt{\frac{24\,M_5^3}{k}} \, e^{-k\,\pi F(x)}$:
\[\mathcal{S}=\dis \frac{2 M_5^3}{k}\int d^4x\sqrt{-g}\left[1-\frac{k\,\Phi^2}{24\,M_5^3}\right]R 
+ \frac{1}{2}\int d^4x\sqrt{-g}\partial_\mu \Phi\partial^\mu \Phi\]
The vacuum expectation value, $\langle \Phi \rangle$, is obtained by introducing a bulk scalar with the interaction terms on both the brane.
This bulk scalar then develops an effective 4D potential on the brane. The minimum of this potential
can be arranged to yield the required value of $kr_c$ as
\begin{equation}\langle \Phi \rangle=\dis \frac{24\,M_5^3}{k}\,e^{-k\pi r_c}.\label{radion1}\end{equation}
The mass of radion field about the minimum is given by, 
 \begin{equation}m_\Phi \sim\frac{kV_b}{2M_5^{3/2}}e^{-kr_c\pi}\label{radion2}\end{equation}
 where $V_b$ is the vev of the bulk stabilising field on the hidden brane.
It can be noticed that the precise mass of the radion is dependent on the backreaction. 
For small backreaction, the expression above dictates radion mass to be of few hundreds 
 GeV.
 
\subsection*{Radion couplings to SM fields} 
Expanding $\Phi$ about its vev as
\[ \Phi = \langle \Phi\rangle + \varphi, \]
The interactions of the radion with matter on the visible brane
can be written as
\begin{equation}
{\cal L}_{int} = \frac{\varphi}{\langle \Phi\rangle} \, \left( T_\mu^\mu \right)\equiv \frac{\varphi}{\Lambda_\varphi} \,  \left( T_\mu^\mu \right)
\label{eqn:Rint}
\end{equation} 
where $\Lambda_\varphi \equiv \langle \Phi\rangle$, $T_{\mu\nu}$ is the symmetric and gauge invariant 
tree-level energy-momentum tensor, defined by
\[T_{\mu\nu} = \frac{2}{\sqrt{-g}}\frac{\delta \mathcal{S}_{matter}}{\delta g^{\mu\nu}}.\]
Restricting to interactions upto quadratic order in SM fields, the tree-level $T_\mu^\mu $ has form 
\begin{eqnarray}
T_\mu^\mu  &=& \sum_f \left(\frac{3}{2}\partial_\mu(\bar{f}i\gamma^\mu f) - 3\bar{f}i\gamma^\mu\partial_\mu f + 4\,m_f\bar{f} f\right)  \\ \nonumber
&-& \partial_\mu h\partial^\mu h + 2m_h^2 h^2 \\ \nonumber &-& 2m_W^2 W^{+\mu}W^-_\mu 
- m_Z^2 Z^\mu Z_\mu   
\label{eqn:trEMtensor}
\end{eqnarray}
where the sum runs over all fermions $f$ and $h$ represents Higgs boson. Note that the radion-SM coupling are
exactly like the coupling of the Higgs boson, except that the SM vacuum expectation
value $v$ is replaced by $\Lambda_\varphi$. Hence, one expects radion phenomenology
to be very similar to Higgs boson phenomenology. However, due to the trace anomaly
, the gauge bosons have additional interaction term
\begin{equation}
{\cal L}^{gauge}_{int} = \sum_i \frac{\beta(e_i)}{2e_i^3} \, F^{\mu\nu i} F_{\mu\nu}^i\,\varphi
\label{eqn:trAnomaly}
\end{equation}
where $\beta(e_i)$ is the beta function corresponding to the coupling $e_i$
of the gauge field $A_i$. The sum is over all the gauge fields in the SM. 
Because of the anomaly term, the radion has sizable interaction strength with
$\gamma\gamma$ and $gg$ pairs, which are completely absent for the SM Higgs. For the case of $W^+W^-$
and $ZZ$ pairs, contribution due to the anomaly term is negligible compared to the
corresponding terms in Eq.~(\ref{eqn:trEMtensor}).

In addition to the above action, radion-Higgs mixing scenario is also possible, which 
we review in the next section.

\subsection*{Radion-Higgs mixing}
Now we discuss the mixed Higgs-radion scenario. This scenario has been discussed 
by several authors\cite{Giudice:2000av,Csaki:2000zn,Dominici:2002jv}, with similar features, but here we choose to work with the
formalism given in Ref.\cite{Csaki:2000zn,Dominici:2002jv}. 
The results in Ref.\cite{Csaki:2000zn,Dominici:2002jv} agrees with Ref.\cite{Giudice:2000av} in the
limit $\dis\frac{v\,\xi}{\Lambda_{\varphi}}\ll 1$ with $\xi$ being the mixing parameter.

The mixing is induced through the following term
\begin{equation}
 \mathcal{L}_{mix.} = -\xi \sqrt{-g_{i}}R(g_{i})H^{\dagger}H
\end{equation} 
where $H = [0,(v + h)/\sqrt{2}]$ with $v = 246\gev$ and $g^i_{\mu\nu}$ is the induced metric. 
After expanding $\sqrt{-g_{i}}R(g_{i})$ to linear order, we get
\[\mathcal{L}_{mix.}=6\xi\gamma h \Box\varphi + 3 \xi\gamma^2\partial_\mu\varphi\partial^\mu\varphi\]
where $\gamma\equiv v/\Lambda_\varphi$. The first term induces kinetic mixing between Higgs
and radion wheras the second term modifies the kinetic term for the radion. The full lagrangian 
including $\mathcal{L}_{mix.}$ becomes
\barr
{\cal L} &=& \frac{1}{2} \partial^\mu h \, \partial_\mu h 
- \frac{1}{2} m_{h}^2 h^2 + \frac{(1+6\gamma^2\xi)}{2} \partial^\mu \varphi \,
\partial_\mu \varphi \nonumber \\ &-& \frac{1}{2} m_\varphi^2 \varphi^2                                                                                                                                                                                                                                                                                                
 - 6\gamma\xi \, \partial^\mu \varphi \, \partial_\mu h  
 \label{eqn:Lagrangian}
\earr
We first normalize the kinetic term using following transformations:
\beq
h = h' + \frac{6\gamma\xi}{Z}\varphi', \quad \varphi = \frac{\varphi'}{Z} 
\eeq
where $h'$, $\varphi'$ are transformed fields, $Z^2 = 1+6\xi\gamma^2(1-6\xi)$ and 
$Z^2$ must be positive to get real mixing matrix and therby positive kinetic term.
 To diagonalize the mass matrix, the following orthogonal transformations 
 are used:
 \beq
 h'=\cos\theta h_m + \sin\theta \varphi_m, \quad \varphi'=- \sin\theta h_m + \cos\theta \varphi_m 
 \eeq
 such that
 \barr
 h &=& \left(\cos\theta-\frac{6\xi\gamma}{Z}\sin\theta\right) h_m + \left(\sin\theta+\frac{6\xi\gamma}{Z}\cos\theta\right) \varphi_m, \nonumber \\
 \varphi &=& - \frac{\sin\theta}{Z} h_m + \frac{\cos\theta}{Z} \varphi_m 
 \earr
 where mixing angle $\theta$ is given as
 \beq
 \tan 2\theta = \frac{12\gamma\xi Z m_h^2}
{m_\varphi^2 - m_h^2 \left(Z^2 - 36\gamma^2\xi^2\right)}
\label{eqn:mixangle}
 \eeq
 The real mixing angle keeps the radion kinetic term positive. This gives us a constraint on 
 $\xi$:
 \[ \frac{1}{12}\left(1-\sqrt{1+\frac{4}{\gamma^2}}\right)\leq \xi \leq \frac{1}{12}\left(1+\sqrt{1+\frac{4}{\gamma^2}}\right)\]
 The physical mass are given by:
 \barr
 m^2_{\varphi_m} &=& \frac{1}{2Z^2} \left(\Xi
- \sqrt{\Xi^2 - 4Z^2 m^2_\varphi m^2_h} 
\right) \\
 m^2_{h_m} &=& \frac{1}{2Z^2} \left(\Xi
+ \sqrt{\Xi^2 - 4Z^2 m^2_\varphi m^2_h} 
\right) 
\label{eqn:mass1}
 \earr
 where $\Xi = m^2_\varphi + (1+6\gamma^2\xi) m^2_h$ and the sign is chosen so that the radion is lighter. From these formulae, it is clear 
 that $\xi,m_h, m_\varphi,\Lambda_\varphi$ are unknown parameters. For our 
 study, we trade $m_h$ and $m_\varphi$ in terms of the physical masses:
\begin{equation}
\begin{array}{rcl}
m_\varphi^2 & = &\dis \frac{Z^2}{2}\left( M^2 -\sqrt{M^2 - \frac{4(1+6\gamma^2\xi) m_{\varphi_m}^2 m_{h_m}^2}{Z^2}} \right)
\nonumber \\ [2.5ex]
m_h^2 & = & \dis \frac{Z^2}{2}\left(\frac{ M^2 + \sqrt{M^2 - \dis \frac{4(1+6\gamma^2\xi) m_{\varphi_m}^2 m_{h_m}^2}{Z^2}}}{(1+6\gamma^2\xi)} \right)
\end{array}
\end{equation}
where $M^2 = m_{\varphi_m}^2 + m_{h_m}^2$
 Thus, to keep $m_\varphi$ and $m_h$ (Lagrangian parameters) real, we must have
 \[\left(m_{\varphi_m}^2 + m_{h_m}^2\right)^2 > \frac{4(1+6\gamma^2\xi) m_{\varphi_m}^2 m_{h_m}^2}{Z^2}
\label{constraint1}
\]
For convenience, we drop the index and redefine $\varphi_m$ and $h_m$ as $\varphi$ and $h$ respectively.

\begin{table*}[htb]
\begin{center}
\begin{tabular}{|c|c|}
\hline
 Coupling	&	Value \\
\hline
  $c_{\varphi\bar{f}{f}}$ & $\dis \frac{-m_f}{v}\left(s_{\theta}+\frac{6\xi\gamma c_\theta}{Z}+\frac{\gamma c_\theta}{Z}\right)$\\
  $c_{\varphi{\gamma}{\gamma}}$	& $\dis \frac{-1}{v} \left[\left(s_{\theta}+\frac{6\xi\gamma c_\theta}{Z}+\frac{\gamma c_\theta}{Z}\right)\left(F_1(\tau_W)+\frac{4}{3}F_{1/2}(\tau_t)\right)-\gamma(b_2+b_Y)\frac{c_\theta}{Z}\right]$	\\
  $c_{\varphi{g}{g}}$	& $\dis \frac{-1}{v}\left[\left(s_{\theta}+\frac{6\xi\gamma c_\theta}{Z}+\frac{\gamma c_\theta}{Z}\right)\frac{F_{1/2}(\tau_t)}{2}-\gamma b_3\frac{c_\theta}{Z}\right]$	\\
  $c_{\varphi{W^+}{W^-}}$&$\dis \frac{2m_W^2}{v}\left(s_{\theta}+\frac{6\xi\gamma c_\theta}{Z}+\frac{\gamma c_\theta}{Z}\right)$\\
  $c_{\varphi{Z}{Z}}$&$\dis \frac{m_Z^2}{v}\left(s_{\theta}+\frac{6\xi\gamma c_\theta}{Z}+\frac{\gamma c_\theta}{Z}\right)$\\  
  $c_{h\bar{f}{f}}$ & 	$\dis -\frac{m_f}{v}\left(\dis c_\theta-s_\theta\frac{6\xi\gamma}{Z}-s_\theta\frac{\gamma}{Z}\right)$\\
  $c_{h{\gamma}{\gamma}}$	&$\dis \frac{-1}{v}\Big[\left(c_\theta-s_\theta\frac{6\xi\gamma}{Z}-s_\theta\frac{\gamma}{Z}\right)\left(F_1(\tau_W)+\frac{4}{3}F_{1/2}(\tau_t)\right)+\gamma(b_2+b_Y)\frac{s_\theta}{Z}\Big]$	\\
  $c_{h{g}{g}}$	&$\dis \frac{-1}{v}\Big[\left(c_\theta-s_\theta\frac{6\xi\gamma}{Z}-s_\theta\frac{\gamma}{Z}\right)\frac{F_{1/2}(\tau_t)}{2}+\gamma b_3\frac{s_\theta}{Z}\Big]$	\\
  $c_{h{W^+}{W^-}}$&$\dis \frac{2m_W^2}{v}\left(c_\theta-s_\theta\frac{6\xi\gamma}{Z}-s_\theta\frac{\gamma}{Z}\right)$\\
  $c_{h{Z}{Z}}$&$\dis \frac{m_Z^2}{v}\left(c_\theta-s_\theta\frac{6\xi\gamma}{Z}-s_\theta\frac{\gamma}{Z}\right)$\\
  \hline
\end{tabular}
\vskip 2ex
\caption{\em Coupling strength of radion and higgs with on-shell SM particles in the mixed radion-Higgs scenario.}
\label{tab:mix_radion}
\end{center}
\end{table*}

In our work, we assume that $\Lambda_\varphi$ is
 greater than the vev of the SM higgs and $\xi$ is of order unity. Such a restriction is necessary as values greater than
 unity are not phenomenologically safe to consider because a large value can change the geometry itself
 through back-reaction.
 
 Next, we consider the effect of mixing on the coupling of higgs and radion to the 
 various SM fields. We list and compare the coupling strength in the Table.~\ref{tab:mix_radion} where the coupling of radion to a pair of vector bosons also includes 
 the trace anomaly term. 

\section{Theoretical and Experimental Constraints}
With the radion-Higgs mixing discussed above, we constrain the model with the limits emanating from unitarity constraints and electroweak (EW) precision data. 
Similarities between the radion and the Higgs boson are utilised to constrain the model further based on the bounds from Higgs exclusion searches 
and Higgs signal measurements.
\begin{itemize}
 \item \textbf{Unitarity}: It is well known that the Higgs boson plays a very crucial role in restoring the perturbative unitarity 
 of gauge boson scattering in the SM. With presence of another scalar in the theory with couplings similar to those of the Higgs boson, it becomes  
 important to inquire if it ruins the perturbative unitarity of the theory. In fact, a lot of work has been done in this 
 aspect of the radion~\cite{Mahanta:2000fw,Bae:2000pk,Choudhury:2001ke,Han:2001xs} and it can be concluded that the contribution of a light radion with $\Lambda_\varphi
 \in [1:5]\tev$ to the amplitudes of these processes are 
 subleading and hence high energy behaviour is majorly decided by the SM higgs boson.
 
  \begin{figure*}
  \centering
  \includegraphics[scale=0.3]{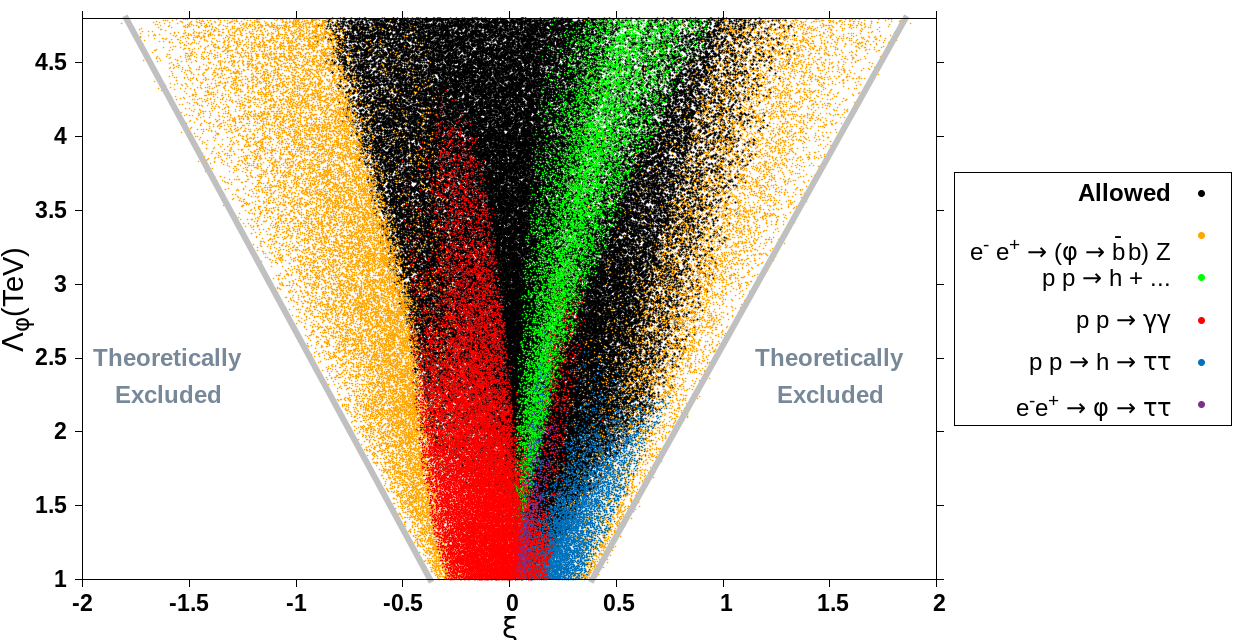}
  \caption{Theoretically and phenomenologically excluded and allowed region in the $\xi-\Lambda_\varphi$ plane obtained from LEP and LHC Higgs exclusion searches.}
  \label{Higgsbounds_exclusion}
 \end{figure*}

 \item \textbf{Electroweak Precision Data}: The oblique parameters can be a useful way to constrain the effects of 
 new physics, especially when the energy scale involved is close to $m_{Z/W}$. Since we consider a BSM scalar in this mass range, it becomes necessary to consider the 
 constraints coming from these measurements~\cite{Csaki:2000zn,Gunion:2003px}. Hence, analysis is made by ensuring that the parameter space satisfies the constraints
 emanating from EW precision measurements.
 
  \item \textbf{Absence of graviton excitations}: Current experimental limits from the LHC rule out any lowest
 graviton excitation of mass below $4.2$ TeV for $k/M_5 \le 0.1$~\cite{Sirunyan:2018exx}. Using Eqn.\ref{radion1} and \ref{radion2},
 this limit translates to a lower bound on $\Lambda_{\varphi}$ of few TeVs. However, this bound can be relaxed considerably
 for models with more than one extra dimensions~\cite{Arun:2015kva,Arun:2016csq,Arun:2016ela}. In these class of models, the mass of graviton and
 its coupling to the SM fields are suppressed due to the presence of two scales in the theory, and, $\Lambda_{\varphi}$ as low as 
 1\,TeV is allowed. In light of this discussion, this analysis has taken $\Lambda_\varphi \geq 1$ TeV.

 \item \textbf{Tevatron, LEP and LHC exclusion limits}: 
 Limits from the non observation of Higgs like resonances at LEP, Tevatron and LHC are imposed on this model through its implementation in  
 HiggsBounds-5.2.0~\cite{Bechtle:2008jh,Bechtle:2013wla,Bechtle:2015pma} and the results are presented in Fig.~\ref{Higgsbounds_exclusion}.
 The major constraint on the parameter space come from the LEP process $e^- e^+\to Z jj/b\bar{b}$. On account of $m_\varphi\,<\,m_h$ as well as 
 enhanced coupling of the $\varphi$ to gluon, $e^- e^+\to Z jj/b\bar{b}$, puts the most stringent bound on the allowed parameter space curtailing
 points at higher $\xi$ for any $\Lambda_\varphi$. The LHC search for the SM Higgs before the Higgs discovery has resulted in the upper limits
 on the scalar production cross section as it restricts $\sigma(p\,p \to h)$ close to the SM values. In this model the Higgs-like particle has an
 enhanced gluon coupling owing to its mixing with the radion in $\xi\,>\,0$ range, leading to enhanced gluon fusion production. Therefore, parts of the parameter space with
 positive $\xi$ are ruled out. Due to the non-minimal contribution of new radion mediated process $p p \to \varphi \to \gamma \gamma$ in the $p p \to \gamma \gamma$,
which is aided by the enhanced gluon fusion production, diphoton channels at the LHC constrain the parameter space of this model.  

\begin{figure*}
 \centering
 \begin{tabular}{cc}
 \includegraphics[scale=0.3]{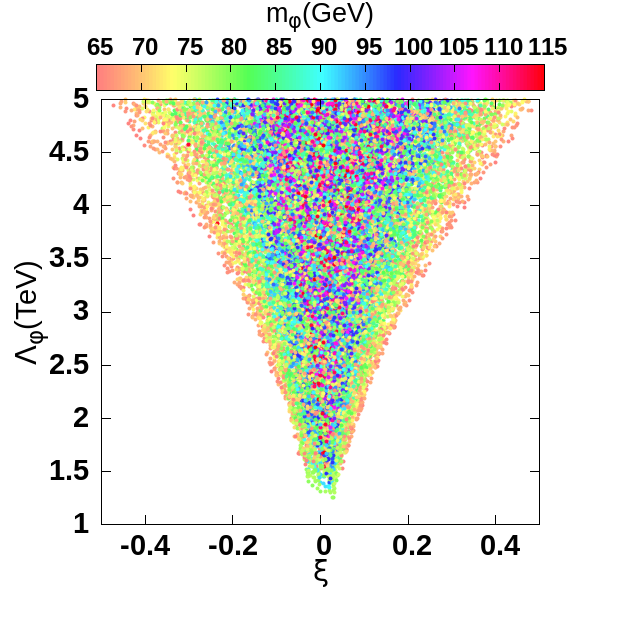}&
 \includegraphics[scale=0.3]{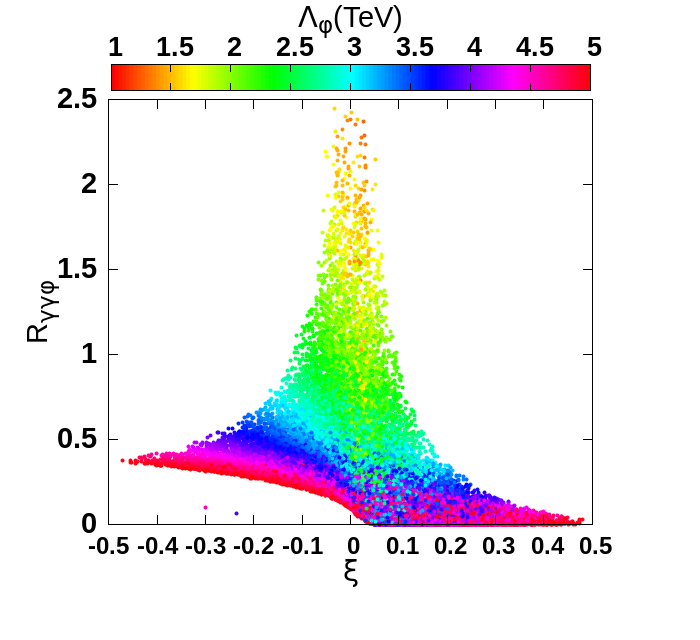}
 \end{tabular}
 \caption{Parameter space allowed from $2\sigma$ measurement of $\mu$-parameter for $\gamma\gamma$, $\bar{b}b$, $\bar{\tau}\tau$, $ZZ$, $W^-W^+$ decay channels {\em (left)} and  
  $R_{\gamma\gamma\varphi}$ values for that parameter space {\em (right)}.}
 \label{HM:2sigma}
\end{figure*}

\item \textbf{Constraints from 125\,GeV Higgs Data}: 
 Next, we analyse the constraints from the Higgs Signal measurements
 which are given in terms of signal strengths ($\mu$ parameters) defined for various decay modes. We have used $\mu$ parameter values 
 for decay modes $\gamma\gamma$, $\bar{b}b$, $\bar{\tau}\tau$, $ZZ$, $W^-W^+$ taken from the LHC data of $\sqrt{s}=13\tev$ and luminosity 39.5 $fb^{-1}$
 ~\cite{Khachatryan:2016vau,Sirunyan:2018koj}. The allowed parameter space for the 2$\sigma$ range of the 
 Higgs Signal measurements is presented in Fig.~\ref{HM:2sigma}~{\em (left)}. Later, we perform the $\chi^2$ test to compare the mixed radion-Higgs
 scenario with the SM. The $\chi^2$ is defined as 
 \begin{equation}
\chi^2 = \sum_{i} \frac{(\mu_{th} - \mu_{i})^2}{\sigma_i^2},
\end{equation} where $\mu_i$ is the $\mu$ parameter quoted above in the i-th channel with 1$\,\sigma$ error bar of $\sigma_i$ and $\mu_{th}$ is the $\mu$ parameter calculated in the model. 
Using this definition the modification in $\chi^2$ compared to that of the SM is given as,
\begin{equation}
\Delta \chi^2 = \chi^2_{SM} - \chi^2_{\varphi h},
\end{equation}
where $\chi^2_{\varphi h}$ is the $\chi^2$ calculated in the mixed radion-Higgs model. 

\begin{figure*}
\centering
\begin{tabular}{ccc}
\includegraphics[scale=0.25]{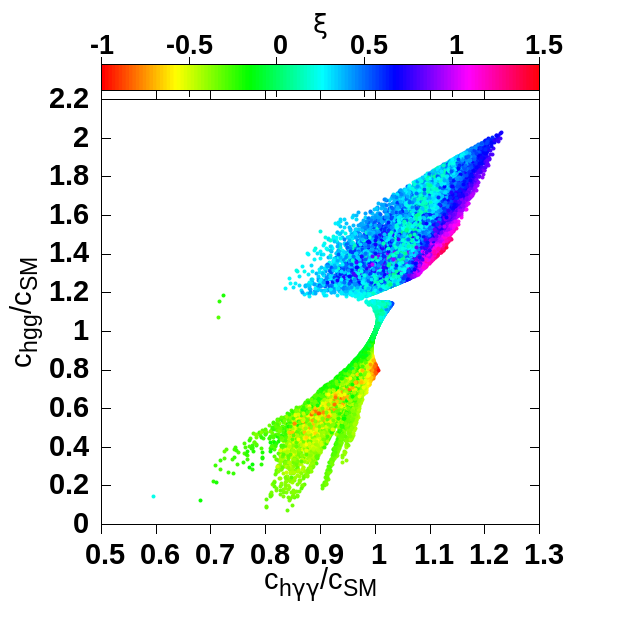}&
\includegraphics[scale=0.25]{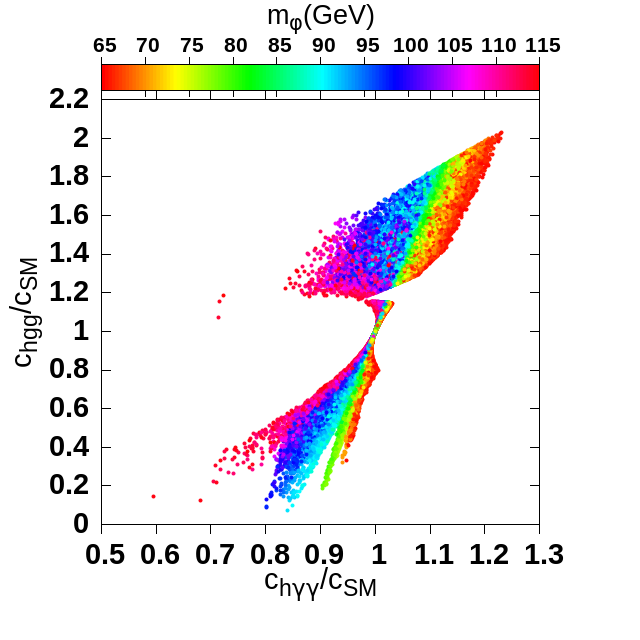}&
\includegraphics[scale=0.25]{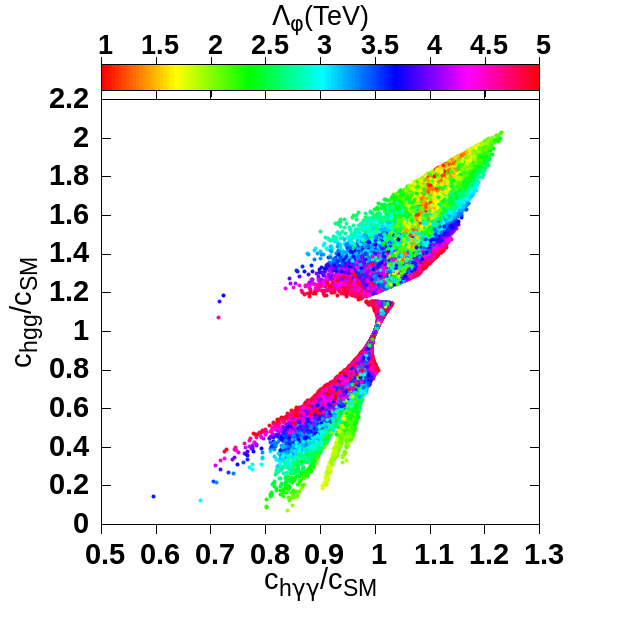}\\
\includegraphics[scale=0.25]{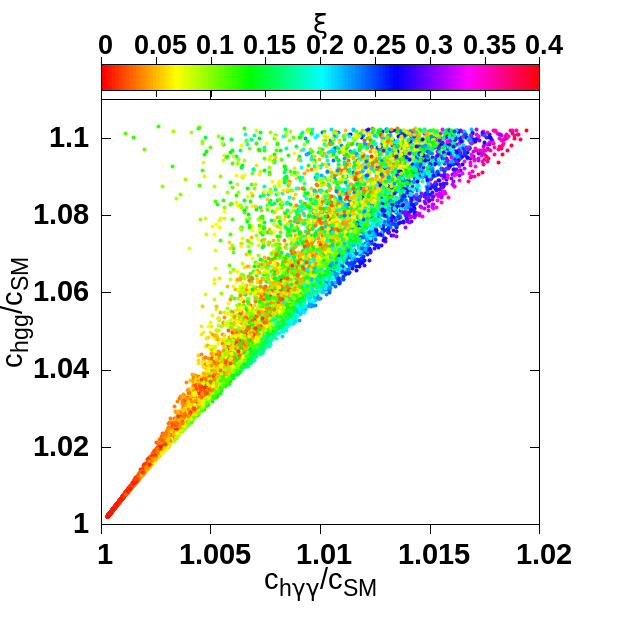}&
\includegraphics[scale=0.25]{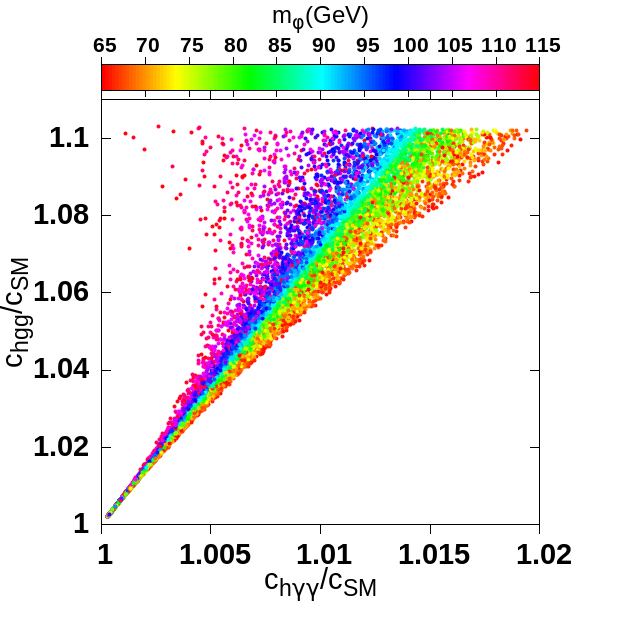}&
\includegraphics[scale=0.25]{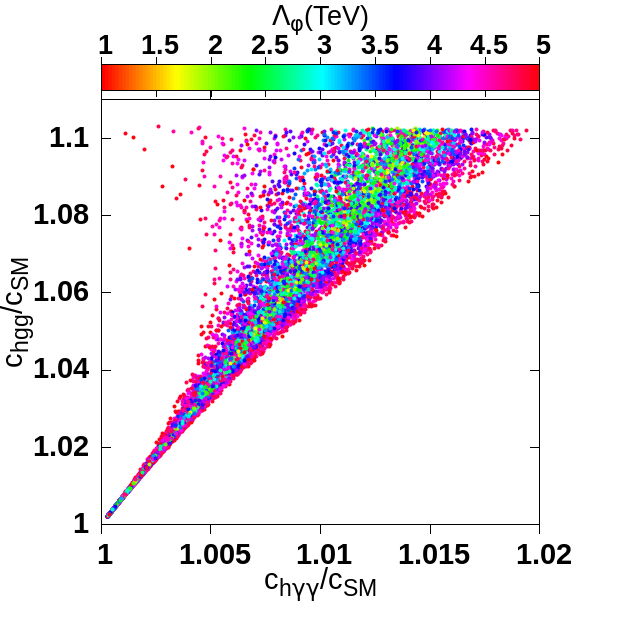}
\end{tabular}
\caption{The parameter space obtained using HiggsBounds(Top) and requiring $\Delta\chi^2\,>\,0$(bottom). $c_{SM}$ is the corresponding 
coupling in the SM.}
\label{correlation}
\end{figure*}

To that end, we project samples on $\xi-\Lambda_\varphi$ and $\xi-m_\varphi$ 
plane for $\Delta \chi^2$ as shown in Fig.\ref{HiggsSignal}. It can be realised that parameter region with $\xi\,>\,0$ which corresponds to $\Delta\chi^2\, > \,0$ fits
the LHC observations better than the SM Higgs. This owes to the fact that positive $\xi$ commensurate with the larger $c_{hii}$ (where $i$ denotes $\gamma,g$) 
and that pushes the $\mu$-parameter values to be greater than one in most of the Higgs decay channels as was quoted above. The correlation between $c_{hgg}$ and $c_{h\gamma\gamma}$ for different parameters before and after 
imposing $\Delta\chi^2\,>\,0$ is compared in Fig.~\ref{correlation}. It should be stressed that, in general, $c_{h\gamma\gamma}$ is more restricted than $c_{hgg}$. 
That is more so in the region where $\Delta\chi^2\,>\,0$ which restricts the $c_{h\gamma\gamma}$ close to the SM values while allowing a 10\% increase for
$c_{hgg}/c_{SM}$ values. This indicates slight increase of Higgs production cross section in gluon fusion channel, giving better fit than the SM.
\end{itemize}

\begin{figure*}
 \centering
 \begin{tabular}{cc}
 \includegraphics[scale=0.3]{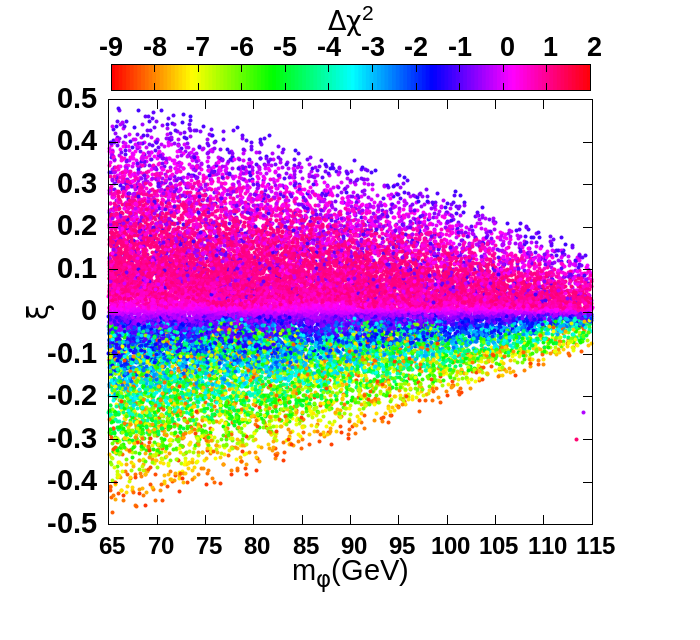}&
 \includegraphics[scale=0.3]{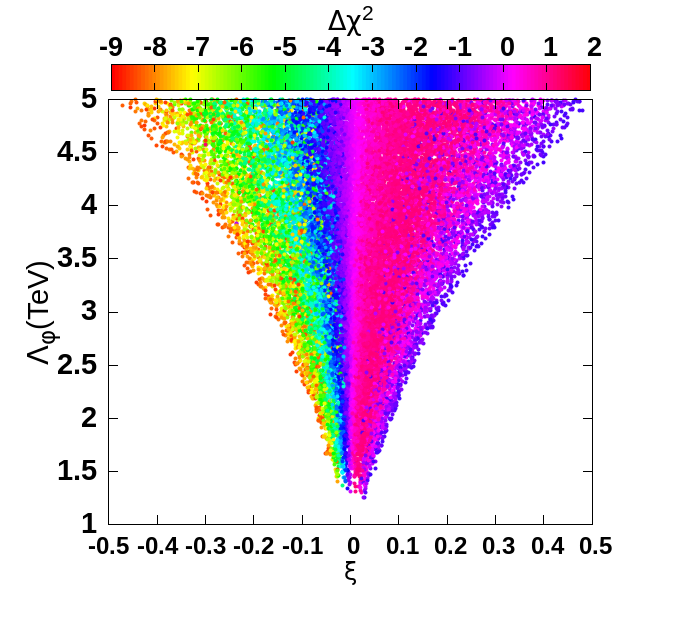}
 \end{tabular}
 \caption{The plot depicts $\xi-\Lambda_\varphi$ and $m_\varphi-\xi$ parameter space for different $\Delta\chi^2$ values. Note that $\Delta\chi^2$ > 0 signifies better fit than that of the SM.}
 \label{HiggsSignal}
\end{figure*}

\section{Results and Discussion}
A recent experimental observation in the light scalar sector is from the CMS~\cite{Sirunyan:2018aui} result that shows a small excess in the diphoton channel near invariant mass of 95 GeV.
This is the most recent in the variety of mild ($\le 3 \sigma$) excesses that have been observed over last few years~\cite{Barate:2003sz,Sirunyan:2017nvi,Sirunyan:2018wim}.
We first invoke radion of the radion-Higgs
mxing model as a light scalar that can potentially give rise to such a diphoton excess at different radion masses and later pin down the parameter space that explains 
the $\gamma \gamma$ channel excess at $m_{\varphi} \approx 95~$GeV. Similar to the SM Higgs, radion is dominantly produced in the gluon fusion mode in this model which is more prominent 
due to enhanced $c_{\varphi gg}$ 
coupling. This enhancement leads to a significant increase in branching ratio (BR) in $\varphi \to g\,g$ mode, reducing the radion branching ratio 
into other final states. However, the drop in BR($\varphi \to \gamma\,\gamma$) is minimal compared to the other channels.


The ratio of radion diphoton signal rate to that of the SM Higgs is defined as
\begin{equation}
R_{\gamma\gamma_{\varphi/h}} = \frac{\sigma(pp \to \varphi) \times \text{BR} (\varphi \to \gamma \gamma)}{[\sigma(pp \to h) \times \text{BR} (h \to \gamma \gamma)]_{SM}}. \nonumber
\end{equation}
With the dominant production mechanism being the gluon fusion for both the SM Higgs and the radion, we rewrite the ratio as
\begin{equation}
R_{\gamma\gamma_{\varphi/h}} \approx \frac{\Gamma( \phi \to gg)}{\Gamma(h \to gg)_{SM}} \times \frac{\text{BR} (\varphi \to \gamma \gamma)}{ \text{BR} (h \to \gamma \gamma)_{SM}}. \nonumber
\end{equation}

\begin{figure*}
\centering
\begin{tabular}{cc}
\includegraphics[scale=0.3]{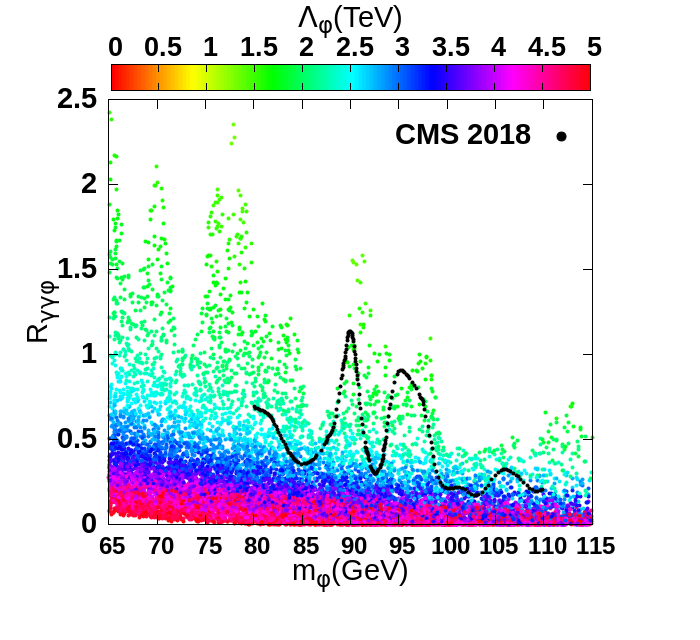}&
\includegraphics[scale=0.3]{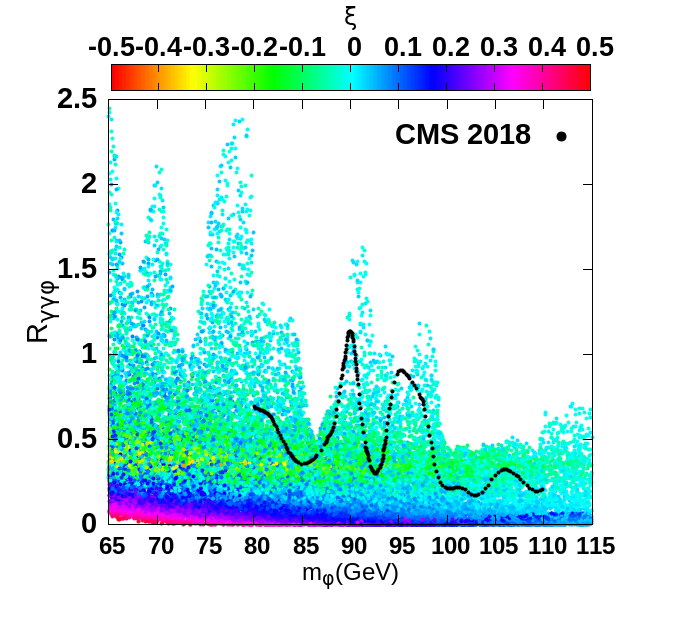}
\end{tabular}
\caption{Allowed parameter space for a better fit of the Higgs signal results than the SM.}
\label{general:lightradion}
\end{figure*}

This ratio depicts the strength of the radion decay to the diphoton channel compared to that of the SM Higgs at respective masses. Here we explore this ratio in
two regions namely where (a) the $\gamma \gamma$ cross section is well below the experimental observations as depicted by the black solid line in Fig.~\ref{general:lightradion}
and (b) there is a significant possibility of observing a $\gamma\gamma$ excess for the points above the black line. A huge part of the parameter space is allowed from
existing constraints such as LHC, LEP, Tevatron exclusion bounds and LHC Higgs signal measurements with the Higgs like scalar fitting the observations better than the
SM Higgs ($\Delta \chi^2 > 0$). Allowed parameter region mainly leads to smaller $R_{\gamma \gamma \varphi}$ and is, therefore, also allowed from light scalar search
at the LHC. This part of the parameter space roughly corresponds
to $\Lambda_\varphi > 2.5~$TeV. 
If we restrict our analysis to the points with $\Delta\chi^2\,>\,0$, the enhancement of radion production
cross section
is significantly curtailed to limit itself to at most twice of the SM number for significant part of the parameter space. Advent of a dominant new decay mode in terms of
$\varphi \to gg$ diminishes the $\varphi \to \gamma \gamma$ branching ratio significantly to restrict $R_{\gamma \gamma\varphi} $ to smaller values. 
For points with smaller positive $\xi$ and smaller $\Lambda_\varphi$ we get some excesses in the diphoton channel for different radion masses. There is an indication 
of excess diphoton events around the 95 GeV radion mass, which was already hinted in CMS observation.
The radion-Higgs mixing predicts towards a hint of radion diphoton rate enhancement at other masses like $m_{\varphi} \sim {70, 77, 90}~$GeV as well. 
Future LHC light scalar search in the diphoton channel will either confirm the presence of radion borne excess or its absence will rule out the parameter space that
shows up the excesses. 

\begin{figure*}
\centering
\begin{tabular}{cc}
\includegraphics[scale=0.3]{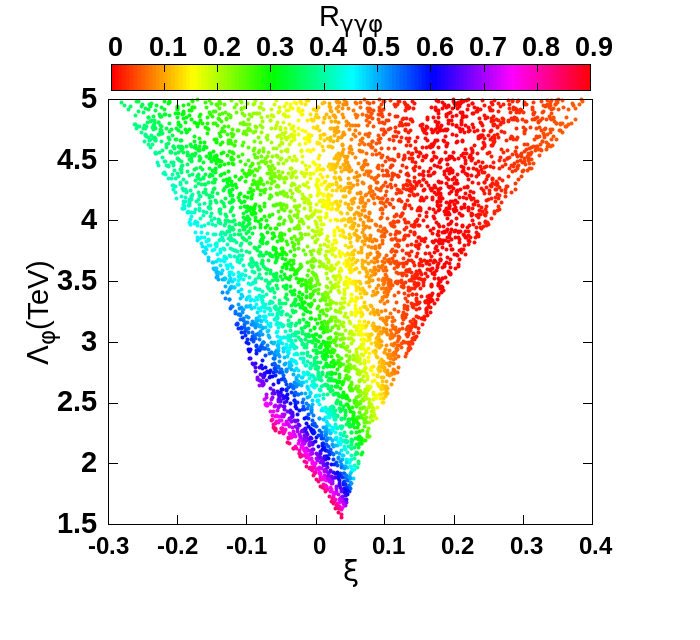}&
\includegraphics[scale=0.3]{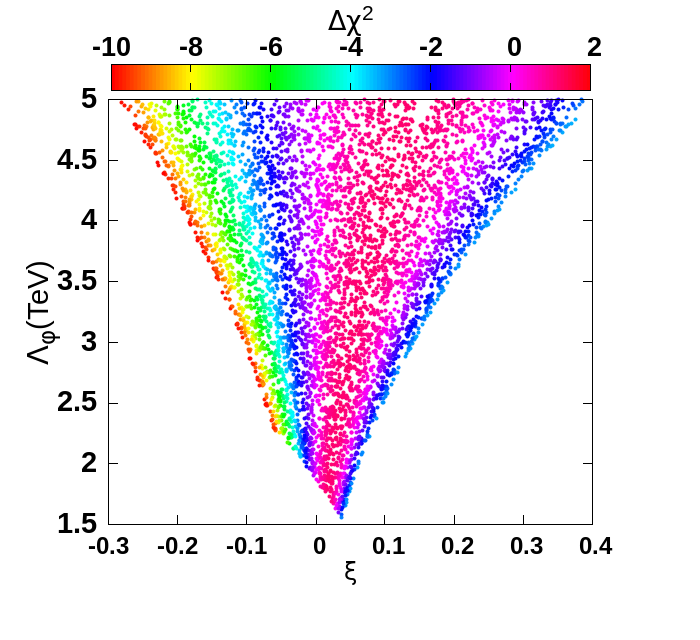}
\end{tabular}
\caption{$\xi-\Lambda_\varphi$ plane for different $\Delta\chi^2$ {\em (left)} and $R_{\gamma\gamma\varphi}$ values {\em (right)} at $m_\varphi\,=\,95\gev$.}
\label{95GeV}
\end{figure*}

We explore the parameter space for $m_\varphi=$ 95 GeV further
in Fig.~\ref{95GeV}. In words, $-0.25\leq\xi\leq 0.35$ and $1.5\tev\leq\Lambda_\varphi\leq5\tev$ satisfies all the theoretical and experimental constraints discussed above.
 For the central value of CMS excess at 95 GeV, $R_{\gamma\gamma\varphi}$ should be around 0.7. Table.~\ref{betterfit_95GeV} shows a few points on $\xi-\Lambda_\varphi$ plane  
 obtained after requiring $\Delta\chi^2 > 0$ and $R_{\gamma\gamma\varphi}\sim 0.7$. 

\begin{table}[ht]
\centering
\begin{tabular}{c|c|c}
 $\xi$ & $\Lambda_\varphi$(TeV) & $\Delta\chi^2$ \\
 \hline\hline
 0.0018 & 2.03 & 0.09  \\
 0.012 & 1.9 & 0.91 \\
 0.024 & 1.84 & 1.05  \\
 0.021 & 1.86 & 1.13 \\
 0.022 & 1.85 & 1.11 
\end{tabular}
\caption{coordinates on $\xi-\Lambda_\varphi$ plane for $\Delta\chi^2 > 0$ and $R_{\gamma\gamma\varphi}\sim $ 0.7 satisfying all theoretical and experimental 
constraints.}
\label{betterfit_95GeV}
\end{table}

To summarise, we have explored the Higgs radion model once again, to discuss the phenomenological prospects of the radion as a
sub-125 GeV BSM scalar and its suitability to produce a diphoton excess at the LHC. Using HiggsBounds, we obtain the parameter space that
is allowed from the Higgs and new scalar search exclusion results from LEP,
Tevatron and LHC together. The parameter space that is ruled out leads us to conclude that $e^+ e^-  \to Z h/\varphi, h/\varphi \to bb/ jj$
is the most constraining for this model, with $p p \to \gamma \gamma$ and $p p  \to h + X$ following the suit to exclude the model points. 
We also report that there is significant parameter space in this model where Higgs like scalar fits better with the LHC data than the SM Higgs. With
positive mixing parameter i.e. $\xi > 0$ and being aided by enhanced gluon fusion production rate, Higgs-like scalar with a tinge of radion is 
a more suitable candidate to be the observed 125\,GeV scalar at the LHC. 

\section*{Acknowledgements}
We would like to thank Debajyoti Choudhury for useful discussions and comments on the manuscript. DS acknowledges the UGC for
financial support. SS thanks UGC for the DS Kothari postdoctoral fellowship grant.

\bibliographystyle{unsrt}
\bibliography{reference}

\end{document}